\documentclass[preprint,12pt]{elsarticle}
\usepackage{ams}
\usepackage{graphicx}

\usepackage{color}

\journal{Journal of Magnetism and Magnetic Materials }

\begin{document}

\begin{frontmatter}




\title{Adaptation and Performance of the Cartesian Coordinates Fast Multipole Method for Nanomagnetic Simulations}


\author{Wen Zhang}
\ead{zhangwen@usc.edu}
\author {Stephan Haas}

\address{Department of Physics and Astronomy, University of
Southern California\\ Los Angeles, CA 90089, USA}

\begin{abstract}
An implementation of the fast multiple method (FMM) is performed for
magnetic systems with long-ranged dipolar interactions. Expansion in
spherical harmonics of the original FMM is replaced by expansion of
polynomials in Cartesian coordinates, which is considerably simpler.
Under open boundary conditions, an expression for multipole moments
of point dipoles in a cell is derived. These make the program
appropriate for nanomagnetic simulations, including magnetic
nanoparticles and ferrofluids. The performance is optimized in terms
of cell size and parameter set (expansion order and opening angle)
and the trade off between computing time and accuracy is
quantitatively studied. A rule of thumb is proposed to decide the
appropriate average number of dipoles in the smallest cells, and an
optimal choice of parameter set is suggested. Finally, the
superiority of Cartesian coordinate FMM is demonstrated by
comparison to spherical harmonics FMM and FFT.
\end{abstract}


\end{frontmatter}

\section{Introduction}
Magnetic nanoparticles and fluids have attracted intensive
investigation during the last two decades
\cite{De'Bell,Liu,bader,fluid1,fluid2}. Micromagnetic simulation is
an powerful tool to study such systems, where the main challenge is
the dipolar interaction between magnetic moments, which can be dealt
with either discretely or in the continuum limit. Since the
interaction is long ranged, the complexity of a brute-force
calculation will be $\mathcal{O}$($N^2$). One popular method to
improve this performance is by Fast Fourier Transform (FFT), which
reduces the complexity to $\mathcal{O}$($N*log(N)$). However, FFT
suffers from the fact that it requires a regular lattice arrangement
of dipoles, and a large number of padding areas have to be added
when dealing with exotic geometries and open boundary conditions.
Thus, the alternative, fast multipole method (FMM) has attracted
increasingly more attention in the past several years
\cite{implm2001,implm2003,psimag,implm2004}.

The FMM was first introduced by L. Greengard $et$ $al$
\cite{greengardbook,greengard}, and has been used ever since to
speed up large scale simulations involving long ranged interactions.
It has the charming advantage of $\mathcal{O}$($N$) complexity.
Furthermore, there is no constraint on the distribution of particles
and the boundaries of the system. Hence it is extremely useful for
magnetic fluid systems and nanomagnets with exotic geometries. Last
but not least, it is a scalable algorithm, i.e. it can be
efficiently implemented in parallel \cite{nakano}. With so many
advantages though, the adoption of FMM in magnetic system is not
widely applied, probably due to two reasons. One concern is that
since there is a huge overhead to achieve $\mathcal{O}$($N$)
complexity, the FMM becomes faster only for very large $N$. Actually
this is true only when one considers a very high order expansion
(say 10). In the context of micromagnetics, we will show that an
expansion to the order of no more than 6 will be satisfying, and
with some optimization procedure FMM will be superior even in a
system of $~10^3$ dipoles. The other reason is that the standard
implementation of the FMM algorithm is based on the well-known
spherical harmonics expansion \cite{jackson} of $1/r$. For dipolar
interaction which decays as $1/r^2$, however, it is not
straightforward to apply. Further, the additional complexity of
calculating spherical harmonics acts as another barrier. To overcome
these shortcomings, expansions in Cartesian coordinate were
proposed\cite{cartesian1, cartesian2}, but they were not applied
successfully until recently\cite{psimag}. Following these studies,
here we present a very simple but useful formula (Eq.
\ref{multipolemoments}) to calculate the multipole moments in
dipolar systems.

In Section II and III, a brief description of the FMM algorithm and
all the necessary equations is presented. In Section IV, performance
issues will be discussed in detail. The number of dipoles contained
in the smallest cell will be shown to be crucial in terms of
performance and a rule of thumb on choosing the right number will be
given (Eq. \ref{so}). Then, the optimal choice of expansion order
and opening angle to achieve certain error bounds will be discussed,
followed by quantitative study of the trade off between computing
time and accuracy. Finally, we compare the Cartesian coordinates FMM
with spherical harmonics FMM and FFT.

\section{Algorithm}

The crucial ideas of the FMM are: (i) chunk source points together
into large cells whose field in remote cells is computed by a
multipole expansion of the source points; (ii) use a single Taylor
expansion to express the smooth field in a given cell contributed by
the multiple expansions of all remote cells. As the details of the
algorithm are discussed in previous literature
\cite{greengardbook,implm2003}, here we will only briefly describe
them, stressing those aspects which are special to our
implementation. Instead of the regular geometric hierarchical
traversals of the system, we follow a simple alternative
way\cite{implm2003} by using recursive functions. This method
simplifies the program significantly and is especially elegant for
system with exotic geometries, since one only needs to recursively
divide the system into halves.

After setting up all the cells, the entire program mainly consists
of three recursive functions traversing the entire system trice.
Pseudocodes are provided in the Appendix.

1. Downward Pass: construct ``partners" list (similar to the
interaction list in the original FMM \cite{greengardbook}). For each
cell, the partners list contain those cells which are near to its
parent but far to itself. Instead of the original geometric rules
(near cell pairs are those cells touching each other), near cell
pairs refer to those with opening angle $\alpha$ larger than certain
value $\alpha_m$ ($\alpha_m<1$). The opening angle is defined as
$2r/d$ ($\alpha_m\equiv 2r/d$), where $r$ is the ``radius" of two
cells and $d$ is the distance between their centers. Note that
``radius" denotes the maximum distance between a corner of one cell
and its center. With these definitions, it is easy to construct the
partners lists. First, set the partners list of each cell to be
empty. Then, for each cell except those cells without children (leaf
cells), start from its partners list generated by its parent (for
the root cell, this list is empty) and perform the following
procedure: (i) add the children of all its near cells it to the
partners list of each of its children, and delete these cells from
its own partners list; (ii) add one of its children to the other
child's partners list (do it for both children); (iii) call its
children to perform these tasks. In this way, we start from the root
cell and recursively traverse its children until reaching leaf
cells. For each leaf cell, add to another list called
``nearPartners" those cells in the partners list which are near to
it and delete them from its partners list. Finally put itself in its
nearPartners list.

2. Upward Pass: construct the multipole moments for each cell. Start
from the leaf cells and explicitly calculate their multipole moments
according to Eq. \ref{multipolemoments} below. Then upward traverse
all the other cells in binary tree hierarchy whose multipole moments
are obtained by shifting the origin of its children's multipole
moments (see Eq. \ref{momentshift}). This process can be easily
realized by a recursive function.

3. Downward Pass: construct the Taylor expansion of the smooth field
in each cell. Start from the root cell and down traverse all the
other cells in the binary tree hierarchy. For each cell, inherit the
Taylor expansion coefficients of the smooth field from its parents
according to Eq. \ref{taylorshift} and then add to it the
contribution from the cells in the partners list according to Eq.
\ref{taylorcalc}.

After these three steps, the only contributions to the magnetostatic
field that are not counted come from dipole pairs among the leaf
cells which are near to each other or within the same cell. So for
each field point in the leaf cell, calculate explicitly Eq.
\ref{H2}, which sum over all the dipoles in those cells in the
nearPartners list except the very dipole located at the field point.

In the majority of implementations of the FMM, the potential is
expanded in terms of spherical harmonics, which complicates the code
and requires long computing time. Previous attempts were made to
expand in Cartesian coordinates \cite{cartesian1,
cartesian2,implm2003}. Initially, the expansion coefficients were
calculated either by hand or in a computationally inefficient way,
but they can in fact be obtained very fast nowadays\cite{implm2003}
using Eqs. \ref{Dn} and \ref{Dnr}. The disadvantage of Cartesian
expansion is that the number of terms below each order is larger
than that with spherical harmonics, but this inefficiency is less
than a factor of two up to order 8. We will show later that in most
cases in the micromagnetic simulation, an order of no more than 6 is
sufficient.

\section{Formalism}

In this section, we provide all the necessary formulae to implement
the above algorithm for nanomagnetic simulation, following Ref.
\cite{implm2003}. To simplify the formalism, the following shorthand
notations are defined:
\begin{eqnarray}\label{def}
\mathbf{n}&\equiv&(n_x,n_y,n_z)\nonumber\\
n &\equiv& n_x+n_y+n_z \nonumber\\
\mathbf{r}^{\mathbf{n}} &\equiv& x^{n_x} y^{n_y} z^{n_z} \nonumber\\
\mathbf{n!} &\equiv& n_x! n_y! n_z!
\end{eqnarray}

Define the multipole moments of a given charge distribution ${q_i}$:

\begin{equation}
Q_\mathbf{n}=\frac{1}{\mathbf{n!}}\sum_i q_i
\mathbf{r}_i^{\mathbf{n}}.
\end{equation}

The evaluation of $Q_n$ is straightforward for monopoles. For
dipoles, however, this is not the case. One way to proceed is divide
the system always into cells which contain only one dipole at the
center, based on the fact that only $Q_{(1,0,0)}$, $Q_{(0,1,0)}$ and
$Q_{(0,0,1)}$ survive for a single dipole $\vec{m}$ located at the
origin, and they equal $m_x$, $m_y$ and $m_z$ respectively.
Nevertheless, this method is not convenient and even inapplicable
for certain systems like ferrofluids. Moreover, having multiple
dipoles in each leaf cell has performance advantages. This issue
will be discussed in detail in Sec IV. Thus it is important to have
an easy formula to calculate multipole moments for dipolar system.
Previous studies have shown complicated formulas for spherical
harmonics expansion \cite{implm2004}. For the multipole moments in
Cartesian coordinate, it becomes much simpler, i.e.

\begin{equation}\label{multipolemoments}
Q_\mathbf{n}=\frac{1}{\mathbf{n!}}\sum_i \vec{m}_i \cdot\nabla
\mathbf{r}_i^{\mathbf{n}}.
\end{equation}

In order to derive this formula, we follow the standard procedure to
treat a dipole $\vec{m}(\vec{r})$ as a limit of two point monopoles
$\pm q$ located at $\vec{r}\pm \vec{d}$ when $d\rightarrow 0$,
keeping $\vec{m}=2q\vec{d}$. After Taylor expansion, Eq.
\ref{multipolemoments} appears.

If $1/|\vec{r}-\vec{r_i}|$ is expanded in Taylor series at $r_i$,
the potential $V(\vec{r})=\sum_i q_i/|\vec{r}-\vec{r_i}|$ can be
expressed in a compact form in terms of the multipole moments,

\begin{eqnarray}\label{Vm}
\frac{1}{|\vec{r}-\vec{r_i}|}&=&
\sum_\mathbf{n}\frac{1}{\mathbf{n}!}
D_{\mathbf{n}}(\vec{r})\mathbf{r}^{\mathbf{n}},\\
 D_{\mathbf{n}}(\vec{r})&\equiv& \frac{\partial^
\mathbf{n}}{\partial
(-\mathbf{r})^\mathbf{n}}(\frac{1}{|\vec{r}|}),\\
 V(\vec{r})&=&\sum_\mathbf{n}
D_{\mathbf{n}}(\vec{r})Q_{\mathbf{n}}.
\end{eqnarray}

In the upward pass (step 2), one needs to shift the origin of the
multipole moments. The moments $Q_{\mathbf{n}}$ about the origin $O$
are related to $Q'_{\mathbf{n}}$ about position $\vec{c}$ (with
respect to $O$) by

\begin{equation}\label{momentshift}
Q_\mathbf{n}=\sum_\mathbf{p}
\frac{\mathbf{c}^\mathbf{p}}{\mathbf{p}!}
Q'_{\mathbf{n}-\mathbf{p}}.
\end{equation}

The Taylor expansion of an arbitrary potential function $V(\vec{r})$
is:

\begin{eqnarray}\label{taylorexpansion}
V(\vec{r})&=& \sum_\mathbf{n}\frac{1}{\mathbf{n}!} V_{\mathbf{n}}
\mathbf{r}^{\mathbf{n}},\\
 V_{\mathbf{n}}&\equiv& \frac{\partial ^\mathbf{n}}{\partial
\mathbf{r}^\mathbf{n}} V(\vec{r})|_{\vec{r}=0}.
\end{eqnarray}

In the second downward pass (step 3), one needs to shift the origin
of the $V_\mathbf{n}$ when a child inherits $V_\mathbf{n}$ from its
parent. The child's $V'_\mathbf{n}$ about position $\vec{c}$ (with
respect to its parent's origin $O$) is related to its parent's
$V_\mathbf{n}$ by

\begin{equation}\label{taylorshift}
V'_\mathbf{n}=\sum_\mathbf{p}
\frac{\mathbf{c}^\mathbf{p}}{\mathbf{p}!} V_{\mathbf{n}+\mathbf{p}}
\end{equation}

Besides, one also needs to calculate the Taylor expansion
coefficient in a given cell from its partner's multipole moments.
This is achieved by

\begin{equation}\label{taylorcalc}
V_\mathbf{n}= (-1)^n \sum_\mathbf{p}
D_{\mathbf{n}+\mathbf{p}}(\vec{c}) Q_{\mathbf{p}},
\end{equation}

\noindent where $\vec{c}$ is the position vector of the center of
this cell with respect to the center of its partner whose multipole
expansion is $Q_{\mathbf{p}}$. It can be shown \cite{implm2003} that
$D_{\mathbf{n}}(\vec{r})$ has the form:

\begin{equation}\label{Dn}
D_\mathbf{n} (\vec{r})=\frac{1}{r^{2n+1}}\sum_{\mathbf{p}(p=n)}
F_\mathbf{n}(\mathbf{p}) \mathbf{r}^\mathbf{p}
\end{equation}

Thus the problem of calculating $D_\mathbf{n}$ breaks down to
creating the table $F_\mathbf{n}(\mathbf{p})$. Note that by
definition of $D_\mathbf{n}$, we have:  can be calculated by
$D_\mathbf{n}$.

\begin{eqnarray}\label{Dnr}
D_\mathbf{n+\hat{x}} (\vec{r}) &=& -\frac{\partial}{\partial x} D_\mathbf{n} \nonumber \\
 &=&\frac{1}{r^{2n+3}} \sum_{\mathbf{p}} F_\mathbf{n}(\mathbf{p}) (-p_x \mathbf{r}^\mathbf{p-\hat{x}+2\hat{y}}\nonumber \\
 &&-p_x\mathbf{r}^\mathbf{p-\hat{x}+2\hat{z}}+(2n+1-p_x)\mathbf{r}^\mathbf{p+\hat{x}})
\end{eqnarray}

\noindent where $\hat{x}=(1,0,0)$. Meanwhile according to Eq.
\ref{Dn},

\begin{equation}
D_\mathbf{n+\hat{x}}
(\vec{r})=\frac{1}{r^{2n+3}}\sum_{\mathbf{p'}(p'=n+\hat{x})}
F_\mathbf{n+\hat{x}}(\mathbf{p'}) \mathbf{r}^\mathbf{p'}
\end{equation}

Similarly one can obtain the such formulas for
$D_\mathbf{n+\hat{y}}$ and $D_\mathbf{n+\hat{z}}$ (a permutation
through x, y, z in above equations). Therefore each $\mathbf{n}$,
$F_\mathbf{n}(\mathbf{p})$ can clearly be built from
$F_\mathbf{n-\hat{\nu}}(\mathbf{p})$ recursively starting from
$F_{(0,0,0)}((0,0,0))=1$, where $\nu$ is chosen among $x$, $y$, $z$
such that $n_\nu>0$. Note that for each $\mathbf{n}$,
$F_\mathbf{n}(\mathbf{p})$ is nonzero only for a few $\mathbf{p}$'s.
Thus it is better to store a list of $\mathbf{p}$s with nonzero
$F_\mathbf{n}(\mathbf{p})$ for each $\mathbf{n}$ and a list of
$F_\mathbf{n}(\mathbf{p})$ value accordingly. Once these are set up,
$D_\mathbf{n}$ can be calculated very fast.

With Eqs. \ref{multipolemoments}, \ref{Vm} and \ref{Dn}, we can now
discuss the criterion to determine near and far cell pairs which is
crucial in the first downward pass (step 1). Suppose we truncate the
terms after the $p_m^{th}$ order. The error is of the order
$r^{p_m}/(d-r')^{p_m}$ \cite{implm2003}, where $r$, $r'$ are the
radius of the source and the field cell and $d$ is the distance
between their centers. In our implementation $r\equiv r'$, thus
$r^{p_m}/(d-r')^{p_m}=[\alpha/(2-\alpha)]^{p_m}$. Obviously, the
series will converge when $\alpha <1$. Thus a maximum $\alpha_m <1$
is chosen to determine whether two cells are near to each other. In
Sec. III, we will discuss how to choose it optimally.


Finally the smooth part of magnetostatic field is given by:
\begin{equation}\label{H}
\vec{H}^s=- \nabla V = \sum_\mathbf{n}\frac{1}{\mathbf{n}!}
V_{\mathbf{n}} \nabla \mathbf{r}^{\mathbf{n}}
\end{equation}

And sum over pairs in the nearPartner list $\mathcal{L}$.
\begin{equation}\label{H2}
\vec{H}_i = \vec{H}^s_i+\sum_{j\in\mathcal{L}, j\neq i}
\frac{\vec{m}_j-3(\vec{m}_i\cdot
\hat{r}_{ij})\hat{r}_{ij}}{r^3_{ij}}
\end{equation}

The total magnetostatic energy $E=-\sum_i \vec{m}_i \cdot \vec{H}_i$
in unit of $\mu_0/4\pi$.

\section{Performance Optimization}

\begin{figure}[h]
\includegraphics[height=.25\textheight]{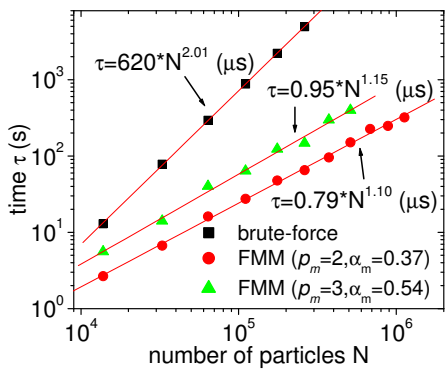}
\caption{(color online) Computational complexity for the benchmark
system. The black square represents the brute-force calculation with
complexity $\mathcal{O}$($n^2$), whereas the red dot and green
triangles stand for the FMM with two different sets of parameters
($\alpha_m$, $p_m$). The red line is a least square fit to the
data.} \label{linear}
\end{figure}

A benchmark where randomly oriented dipoles are arranged on a simple
cubic lattice was adopted to test the performance of our program.
All the results below are averaged over 50 random configurations and
performed on an dual Intel P4 3.0GHz with 2G RAM. Fig. \ref{linear}
demonstrated that the complexity of our program is $\mathcal{O}(n)$,
and it begins to outperform the brute-force calculation around
$N_c=1000$. The small deviations from linearity are caused by the
change of hierarchy as the size of the systems increases. The two
parallel lines with different colors and symbols correspond to two
sets of $\alpha_m$ and $p_m$ achieving the same accuracy. The shift
between them indicates that there exists an optimal choice of
$\alpha_m$ and $p_m$ set for a given accuracy. In the following
analysis, we choose a $32*32*32$ cubic system, but the conclusions
are independent of system size unless it is too small.

Before discussing of the parameter set, there is another important
issue which affects the performance significantly and needs to be
elucidated first. It is the smallest cell size or the average number
of particles ($s$) in the smallest cell. The value of $s$ determines
the number of levels in the binary tree and how many brute-force
calculations are required on the finest level. Choosing an
inappropriate $s$ can deteriorate the performance by a factor of 10.
Thus it is indeed a nontrivial issue. Obviously, the larger $s$ is,
the fewer levels and the more brute-force calculations will be
required. Because of the fairly large overhead of FMM with respect
to the brute-force calculation, it is naturally expected that $s=1$
will not be a good solution. And it is also not good to make $s$ too
large, for we will lose the power of the FMM then. So there must be
an optimal number $s_o$ which gives best performance for a given
parameter set. Meanwhile, since the overhead increases as the
expansion order $p_m$ increases, the optimal $s_o$ is expected to be
a function of $p_m$ and should increase with $p_m$ as well. On the
other hand, $\alpha_m$ determines the number of cells in the ``near"
region of each specific cell, but the ratio of the number of cells
in the partners list and the number of cells in the nearPartners
list should be independent of it. If this ratio remains constant,
there is no reason to change $s_o$ for fixed $p_m$. Therefore, $s_o
$ only depends on $p_m$. Though it is not too difficult to come to
this conclusion, it is still not clear at all how to choose $s_o$.
Here, we offer a solution. Table I shows results for a $32^3$ system
with $p_m=3$, $\alpha_m=0.54$. In this case, $s_o=8$. However as for
arbitrary system sizes, we cannot divide them freely, the dependence
of $s_o$ on $p_m$ is a discreet function, and different for
different system sizes. We overcome this problem by averaging $s_o$
over 30 system size from $30^3$ to $59^3$. The result is shown in
Fig. {\ref{cellsize}, where each number beside the square points is
$s_o$ for $p_m$ from 2 to 12. The line is a linear fit for those
points with $p_m>3$, indicating an approximate power law
relationship between $s_o$ and $p_m$ and the power is approximately
2. As $s_o$ for a given system size will never be the value
indicated in the graph anyhow, we reach an empirical equation as
follows:

\begin{equation}\label{so}
s_o =\{
\begin{array}{c}
 0.74p_m^2 \qquad  \ p_m>3  \\
\qquad 8 \qquad \qquad \ otherwise
\end{array}
\end{equation}

It is very easy to use this equation to estimate $s_o$. Lastly, for
the same $s_o$, one can have different divisions of the system, but
it is always better to have the smallest cell as close to a cube as
possible.

\begin{table}[ht]\label{32}
\caption{Performance for ($p_m=3$, $\alpha_m=0.54$)} 
\centering 
\begin{tabular}{ c @{\qquad    } c  }
  \hline\hline
   $s$& $time$\\ \hline
    1 &    34.82\\
    2 &    14.69 \\
    4 &      8.28 \\
    8 &   6.58\\
    16 &     9.13 \\
    32 &13.69\\
    64 & 20.58\\

  \hline

\end{tabular}
\end{table}

\begin{figure}[h]
\includegraphics[height=.25\textheight]{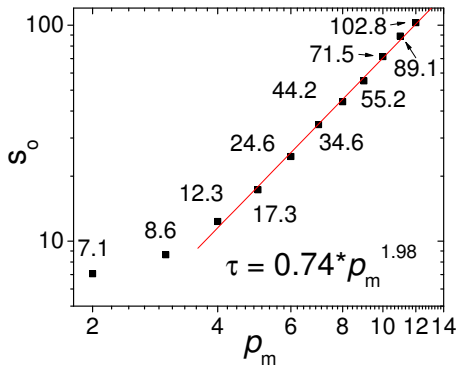}
\caption{(color online) The optimal number of particles in the
smallest cell $s_o$ as a function of expansion order $p_m$. The red
line is fitted by the data ($p_m>3$). The number next to each data
point is the actual $s_o$.}\label{cellsize}
\end{figure}

Based on above discussion, we propose a rule of thumb on how to
choose the smallest cells: (i) calculate $s_o$ according to Eq.
\ref{so}; (ii) the smallest cells do not have to be cubic and should
contain as close to $s_o$ particles as possible; (iii) under the
above two conditions, the smallest cells should be as close to cubic
as possible. For example: (i) for $50^3$ system with $p_m=3$, the
best smallest cell choice is 1.5625*1.5625*3.125 where each cell on
average contains 7.63 particles; (ii) for 40*40*10 with $p_m=4$, the
best choice is 2.5*2.5*2.5 ($\sim$15.6); (iii) for $32^3$ with
$p_m=5$, the best choice is 2*2*4($\sim$16).

\begin{table}[ht]\label{e}
\caption{Performance for $e=0.05$} 
\centering 
\begin{tabular}{ c @{\qquad    } c @{ \qquad    } c }
  \hline\hline
   $p_m$& $\alpha_m$ & time\\ \hline
    2 &  0.36   &   14.15\\
    3 &  0.54 &   6.71 \\
    4 &  0.64 &   6.80\\
    5 &  0.71 &   6.22\\
    6 &  0.76 &   7.36\\
    7 &0.79 & 8.14\\
    8 & 0.81 &10.65\\
  \hline

\end{tabular}
\end{table}

\begin{figure}[h]
\includegraphics[height=.25\textheight]{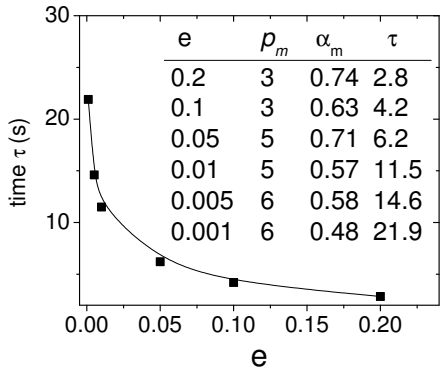}
\caption{Optimal computing time versus accuracy. The curve is a
guide for the eye. The inset table shows the optimal choice of
parameter set($p_m$, $\alpha_m$) and the corresponding computing
time for each error $e$.}\label{time}
\end{figure}

Now that we know how to divide the system, we can optimize the
parameter set ($p_m$, $\alpha_m$). Use
$e\equiv[\alpha_m/(2-\alpha_m)]^{p_m}$ as the error estimate.
Obviously, the error decreases as $p_m$ increases and $\alpha_m$
decreases. To achieve certain accuracy, one can have various sets of
($p_m$, $\alpha_m$) combinations. The computational effort, on the
other hand, increases as $p_m$ increases and $\alpha_m$ decreases.
This claim is easy to understand for $p_m$, since larger $p_m$
results in more terms in the expansion. $\alpha_m$, on the other
hand, determines how many cells there will be in the partners list
and nearPartner list. The smaller it is, the more cells there are in
the lists, resulting in more computing effort. Therefore the
computing effort varies for different sets of ($p_m$, $\alpha_m$).
Table II shows one example for $e=0.05$. Obviously, the choice of
($p_m=5$, $\alpha_m=0.71$) is optimal in this case. The inset table
in Fig. \ref{time} gives the optimal choices of ($p_m$, $\alpha_m$)
for different accuracies. The curve shows the optimal computing time
as a function of accuracy. The computing time grows exponentially
when $e<0.05$, so it suggests to choose $e=0.05$ if the error is
acceptable. In fact, the error of magnetostatic energy for $e=0.05$
is less than $1\%$ except for configurations whose energy are close
to zero. This already reaches the requirement of micromagnetics
where other uncertainties may make it pointless to go beyond such
accuracy. Thus ($p_m=5$, $\alpha_m=0.71$) is suggested, if no other
requirements exist. Even if higher accuracies are required,
($p_m=6$, $\alpha_m=0.48$) will definitely be sufficient, since the
average percentage error in this case will be $0.02\%$. Thus, the
inefficiency caused by Cartesian coordinates is always less than 2.


\begin{figure}[h]
\includegraphics[height=.25\textheight]{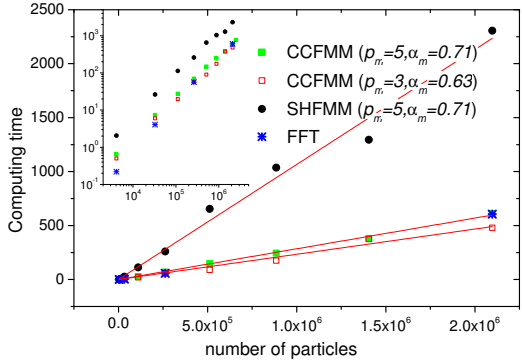}
\caption{(color online) Comparison of performance of Cartesian
coordinate FMM (CCFMM) (green solid square: $p_m=5,\alpha_m=0.71$;
red open square: $p_m=3,\alpha_m=0.63$), spherical harmonics FMM
(SHFMM)(black dot: $p_m=5,\alpha_m=0.71$) and FFT method (blue
star). The inset is a log scale plot. Using Cartesian coordinates
makes the FMM roughly three times faster. With
$p_m=5,\alpha_m=0.71$, the performance of CCFMM is comparable to the
FFT method. When the system is sufficiently large or lower accuracy
is satisfactory, CCFMM can be superior to FFT.} \label{comp}
\end{figure}

Finally we compare the performance of the Cartesian coordinate FMM
(CCFMM) with that of the spherical harmonics FMM (SHFMM) and the FFT
method. For SHFMM, we use the same tree hierarchy as our CCFMM so
that all the performance difference attributes to the difference of
base function in expansion: CCFMM uses simple polynomial expansion,
while SHFMM uses spherical harmonics, which is much more
computationally expensive. Regarding the error boundary, these two
methods is very similar as long as we keep the parameter set $\{s,
p_m, \alpha_m\}$ the same. They will only differ by a small factor
of order  1. As to FFT, the routine given in $\textit{Numerical
Recipes}$ \cite{nr} is used. The results are summarized in Fig.
\ref{comp}. As expected, the computing time increases linearly for
both FMM, and the FFT method gives a typical
$\mathcal{O}$($N*log(N)$) scaling behavior, which can be seen more
clearly on the log scale plot. Focusing on the two FMM, CCFMM is
clearly superior to SHFMM. This is all due to the expensive
calculation of trigonometric functions of spherical harmonics.
Further, cumbersome complex number calculations are avoided by
working in Cartesian coordinates. The gain is approximately a factor
of three given the chosen parameters. This result is comparable to
what has been reported in previous literature\cite{psimag}. However,
differences arise because we use a higher expansion order ($p_m=5$)
while earlier authors used ($p_m=4$). As a rule of thumb: the higher
the order, the less favorable the cartesian expansion. Another
factor that results in different speed-ups is the method used to
calculate spherical harmonics. In our case, we hard-code the
expression of spherical harmonics rather than using recursion
relations. Meanwhile we have also minimized the trigonometric
functions evaluations. With these optimization of SHFMM though, we
found the CCFMM clearly exhibits superior performance at relatively
lower expansion ($p_m<8$). To achieve a decent $1\%$ average error,
parameter set $\{p_m=5,\alpha_m=0.71\}$ is recommended. In this
case, the performances of FFT and CCFMM are similar. Both the
benefit and limitation of FFT is clear. It is an exact method, and
it is easy to implement. However, FFT requires a uniform simple
cubit grid and a large padding area for exotic structures with open
boundary condition. Even though special FFTs can be adapted to
nonuniform systems, they become problem specific and complicated.
FMM, on the other hand, is quite generic and natural for nonuniform
systems and exotic structures with open boundary condition. What's
more, it can beat FFT when only lower accuracy is needed, as shown
by the red open square in Fig.\ref{comp}.


\section{Conclusions}
In conclusion, we have implemented and analyzed the performance of
the Cartesian coordinate based FMM in dipolar systems. A brief
description of the algorithm and supplementary equations are
provided. In particular, we present a simple formula (Eq.
\ref{multipolemoments}) to calculate the multipole moments for point
dipoles. While carefully analyzing the performance by comparing with
spherical harmonics FMM and FFT, we have shown that the Cartesian
coordinate based FMM is appropriate for magnetic simulations due to
their moderate accuracy requirements. A rule of thumb to decide the
optimal number of dipoles in the smallest cell is proposed (Eq.
\ref{so}), and the optimal choices of the expansion order and
opening angle under different precision requirements are discussed.
($p_m=5$, $\alpha_m=0.71$) is suggested for nanomagnetic simulation
generally. As the FMM is a parallel scalable algorithm, it can be
efficiently implemented in parallel architectures.\\


\textsc{\textbf{Acknowledgement:}} We would like to thank Aiichiro
Nakano and Yaqi Tao for useful discussions. The computing facility
was generously provided by the University of Southern California
high-performance supercomputing center. We also acknowledge
financial support by the Department of Energy under grant
DE-FG02-05ER46240.\\

\appendix
\section{APPENDIX}
Here we provide the pseudocode for the three recursive function
mentioned in Sec. II. Within the framework of object-oriented
programming, all these function are public members of an object
called ``cell".\\

\noindent  void cell::createPartners()

\hangafter 0 \hangindent 1em \noindent FOR each cell ``A" in the
``Partners" list

\hangafter 0 \hangindent 2em \noindent IF the cell A is near to this
cell THEN

\hangafter 0 \hangindent 3em \noindent  IF this cell has child THEN

\hangafter 0 \hangindent 4em  \noindent  put the children of A into
the ``Partners" list of the children of this cell;\\erase A from the
Partners list of this cell;

\hangafter 0 \hangindent 3em \noindent ELSE

\hangafter 0 \hangindent 4em \noindent Put A into the NearPartners
list of this cell;\\erase A from the Partners list of this cell;

\hangafter 0 \hangindent 3em \noindent END IF

\hangafter 0 \hangindent 2em \noindent END IF

\hangafter 0 \hangindent 1em \noindent END FOR

\hangafter 0 \hangindent 1em \noindent IF this cell has child THEN

\hangafter 0 \hangindent 2em \noindent put one of its children into
the ``Partners" list of the other children (do this for both
children);\\call its children's CreatePartners();

\hangafter 0 \hangindent 1em \noindent ELSE

\hangafter 0 \hangindent 2em \noindent Add this cell into the
NearPartners list of itself;

\hangafter 0 \hangindent 1em \noindent END IF

\noindent  END FUNCTION\\

\noindent void cell::updateMoment()

\hangafter 0 \hangindent 1em \noindent   clear the multipole moments
of this cell;

\hangafter 0 \hangindent 1em \noindent   IF this cell is not the
smallest cell  Then

\hangafter 0 \hangindent 2em \noindent        call its children's
updateMoment();

\hangafter 0 \hangindent 2em \noindent        sum over its
children's multipole moment with shift of origin;

\hangafter 0 \hangindent 1em \noindent ELSE IF there is dipoles in
this smallest cell THEN

\hangafter 0 \hangindent 2em \noindent   calculate the multipole
moments of this cell directly from the dipole distributions;

\hangafter 0 \hangindent 1em \noindent  END IF

\noindent END FUNCTION\\

\noindent double cell::updateField()

\hangafter 0 \hangindent 1em \noindent  IF this cell has parent THEN

\hangafter 0 \hangindent 2em \noindent inherit the Taylor expansion
from its parent with shift of origin;

\hangafter 0 \hangindent 1em \noindent END IF

\hangafter 0 \hangindent 1em \noindent FOR each cell in the
``Partners" list

\hangafter 0 \hangindent 2em \noindent add to the Taylor expansion
of this cell the field generated by A;

\hangafter 0 \hangindent 1em \noindent END FOR

\hangafter 0 \hangindent 1em \noindent IF this cell has child THEN

\hangafter 0 \hangindent 2em \noindent call its children's
updateField();\\ calculate the total energy of this cell by the sum
of the energy of its children;

\hangafter 0 \hangindent 1em \noindent ELSE

\hangafter 0 \hangindent 2em \noindent calculate smooth local field
from the Taylor expansion contributed from all far cells;\\FOR each
cell ``A"  in the ``nearPartners" list

\hangafter 0 \hangindent 3em \noindent add to the local field from
each dipoles in A except when the dipole is located at the field
point;

\hangafter 0 \hangindent 2em \noindent END FOR

\hangafter 0 \hangindent 2em \noindent calculate the energy of this
cell by summing over all the dipole in it and divide the energy by 2
to eliminate double inclusion of the interaction energy

\hangafter 0 \hangindent 1em \noindent END IF

\hangafter 0 \hangindent 1em \noindent return the energy of this
cell;

\noindent END FUNCTION

\bibliographystyle{elsarticle-num}
\bibliography{scale}        

\end{document}